\documentclass[acmlarge]{acmart} 

\AtBeginDocument{%
  \providecommand\BibTeX{{%
    \normalfont B\kern-0.5em{\scshape i\kern-0.25em b}\kern-0.8em\TeX}}}

\begin{document}

\title{Pattern Discovery in Time Series  with Byte Pair Encoding}

\author{Nazgol Tavabi}
\affiliation{%
  \institution{USC Information Sciences Institute}
  \country{USA}
}
\email{nazgolta@isi.edu}

\author{Kristina Lerman}
\affiliation{%
  \institution{USC Information Sciences Institute}
  \country{USA}
}
\email{lerman@isi.edu}







\renewcommand{\shortauthors}{Tavabi et al.}

\begin{abstract}
The growing popularity of wearable sensors has generated large quantities of temporal physiological and activity data. Ability to analyze this data offers new opportunities for real-time health monitoring and forecasting. However, temporal physiological data presents many analytic challenges: the  data is noisy, contains many missing values, and each series has a different length.  Most methods proposed for time series analysis and classification do not handle datasets with these characteristics nor do they offer interpretability and explainability, a critical requirement in the health domain. We propose an unsupervised method for learning representations of time series based on common patterns identified within them. The patterns are, interpretable, variable in length, and extracted using Byte Pair Encoding compression technique. In this way the method can capture both long-term and short-term dependencies present in the data. We show that this method applies to both univariate and multivariate time series and beats state-of-the-art approaches on a real world dataset collected from wearable sensors.
\end{abstract}




\maketitle

\section{Introduction}
Rapid technological advances have enabled continuous collection of physiological and activity data from wearable sensors in the form of time series. Such data offers new opportunities to quantify and characterize human behavior, monitor health, and assess psychological well-being in real-time \cite{banaee2013data}.  

However when it comes to modeling multiple time series there are many different variants of this problem, each with its own set of challenges. These variants include time series with, different lengths, multiple modalities--multivariate time series--, missing data, noise, etc. Because of these challenges, many methods put constraints on the input data and can be used only on a subset of problems. Data collected from wearable sensors, usually has all of the characteristics mentioned above, hence few methods can be applied to it.

Explainability and interpretability are very helpful and important properties of machine learning approaches. Having interpretable features is necessary in applications like healthcare, where doctors and physicians need to understand the reasoning behind a decision made by the model. However, most of the state-of-the-art methods lack these properties, especially methods that rely on neural networks and are mostly black box models.

In this work we propose an unsupervised, explainable and interpretable method for learning representations from time series. This method, is scalable, can handle different variants of time series datasets, has low computation complexity, and beats state-of-the-art methods on a real world dataset. Our proposed approach extracts common patterns from the data by discretizing the time series and using Byte Pair Encoding compression technique (BPE). Afterwards, series in the dataset are represented by observed frequencies of identified patterns. The identified patterns are interpretable and can be of different lengths. Thus, the method can capture both long term and short term dependencies present in the data.

\section{Related Work}
Many methods have been proposed  for time series classification. \cite{bagnall2017great} covers a wide variety of them and categorizes time series classification methods to six groups. Based on the recent advancement of neural network and deep learning methods, we see fit to add a seventh category for methods using deep learning approaches for time series classification, \cite{fawaz2019deep} reviews successful deep learning approaches for time series data. The seven categories are listed below.   

 \textit{Whole series} In these algorithms a similarity measure is defined between time series. 
The most notable method in this group is Dynamic Time Warping (DTW) \cite{muller2007dynamic} with Nearest Neighbour classifier, which is still used as a baseline for many methods. 

 \textit{Intervals} These types of algorithms extract features from intervals, instead of the whole time series. For example in a long time series, an interval can be considered a minute/hour, and simple features such as mean and standard deviation can be calculated from each interval. \cite{burghardt2020having} is a recent example and \cite{rodriguez2004support} is one of the earliest works with this approach.
 
 \textit{Shapelets} In these methods the focus is to find short patterns, commonly called shapelets, in time series which can distinguish between different classes. In these algorithms the presence, or lack thereof the shapelet in the series is important and its location is irrelevant \cite{hills2014classification,bostrom2015binary}. Most shapelet algorithms enumerate identified shapelets and choose the ones that help with the classification \cite{ye2011time,rakthanmanon2013fast}. The alternative approach is to capture different shapelets in an unsupervised manner, then perform classification on the generated features \cite{hills2014classification,bostrom2015binary}. The latter approach can help reduce overfitting compared to the former. Also unsupervised approaches can be easily extended to regression tasks while it may not be feasible to extend a supervised classification task like \cite{ye2011time} to a regression problem. 

 \textit{Dictionary based} Our proposed approach fits in this group. This type can be seen as an extension of Shapelet methods. Instead of forming the decision boundaries based on presence of patterns, features generated by dictionary based methods, capture the relative frequency of observed shapelets in each series. Most of the methods in this group first discretize the series and convert them to strings of symbols, then identify the patterns. 
 One of the most popular methods for discretizing time series is Symbolic Aggregate Approximation (SAX) \cite{lin2007experiencing}. In SAX, the normalized series are first transformed by Piecewise Aggregate Approximation (PAA) \cite{keogh2001dimensionality}, then discretized into bins formed from equal probability areas of the normal distribution. PAA reduces the dimension of the input time series by splitting them into segments and averaging the values in these segments. 
In \cite{lin2012rotation} after SAX transformation, signals are broken down to windows with a fixed size L which are referred to as words, and a histogram of word counts is used as features for classification similar to bag-of-words approach. \cite{senin2013sax} has a similar approach but instead of bag-of-words it uses TF-IDF (Term Frequency-Inverse Document Frequency), and instead of calculating TF-IDF for each time series, it calculates them for each class. 
Probably the most popular methods in this group is Bag-of-SFA-Symbols (BOSS) model \cite{schafer2015boss}, which similar to DTW, is used as a baseline in many recent methods. Instead of using PAA, BOSS applies Discrete Fourier Transform (DFT) on each window; 
then discretizes the series. Most of the methods in this group, when instances of consecutive identical words such as ``aba aba`` in ``aba aba acc`` are observed,  only count the first word and ignore the remaining repeated words to avoid over counting trivial matches. 

 \textit{Combinations} This group contains methods that combine elements or features from different approaches.
Collection Of Transformation Ensembles (COTE) \cite{bagnall2015time} is known as one of the most accurate methods for time series classification. COTE combines different classifiers over representations of data in different domains including time and shapelet. HIVE-COTE \cite{lines2018time}, an extension to COTE, proposes a hierarchical structure with probabilistic voting over multiple classifiers and has also proven to be very successful.  

 \textit{Model based} In these methods, generative models are fitted to time series and the similarity between the series is measured by comparing the similarity between the models fitted to them. For example \cite{kalpakis2001distance} measures the distance between series by comparing their ARIMA models and
 \cite{tavabi2019characterizing,tavabi2020learning}  compare the non-parametric Hidden Markov Models fitted to each series. 

 \textit{Deep learning} 
 Although intuitively, it seems that recurrent neural networks might be better suited for time series classification \cite{malhotra2017timenet,husken2003recurrent}, 
 convolutional networks are proven to be more successful in this task \cite{zhao2017convolutional}. An important work in this group is ROCKET (RandOm Convolutional KErnel Transform) \cite{dempster2019rocket}. ROCKET transforms time series with a large number of random convolutional kernels, i.e., kernels with random lengths, weights, biases, etc. It has become popular for its exceptional computation speed and accurate results. 
 
 Another notable work is an unsupervised embedding approach proposed in \cite{franceschi2019unsupervised}. These embeddings are generated by an encoder with causal dilated convolutional layers. If we annotate the time series in the dataset as $x_1 \cdots x_N$; the model is trained such that the embeddings of sub-series in $x_i$ 
are closer to each other than to a sub-series in $\forall_{j\ne i}x_j$ . Both of these methods generate unsupervised representations, and are able to handle different variants of time series (variable length, multivariate, etc). They are also among the best and most recent methods proposed for time series classification. We compare the results of our approach with these two methods.

It should be mentioned that although time series classification has gained more attention, there are also many time series regression problems such as affect estimation from wearable sensors \cite{yan2020affect}, analysis of fMRI signals \cite{formisano2008multivariate}, and many more examples. Many of the methods described above, such as methods in the \textit{Combinations} category, can't be extended for a regression problem. Also, most methods in \textit{Deep Learning} category are not interpretable nor explainable. 
\textit{Dictionary Based} methods are known for being interpretable, but in most cases the identified patterns have the same length. 
There are a few methods in motif discovery for identifying patterns with variable length \cite{torkamani2017survey}. However, since these methods were proposed for identifying the best set of motifs, they can't be easily compared with dictionary based methods. Motif discovery methods are mostly evaluated based on their scalability, robustness to noise, exactness, and other related characteristics and are not necessarily targeted for classification/regression tasks. 

As previously mentioned, a standard practice in dictionary based methods is that, when instances of consecutive identical words are observed, the first word is counted and the remaining repeated words are ignored. However, we argue that removing data, even redundant data, can lead to information loss. In this approach, time series data is first transformed into different variations: in one variation the repeated words are kept, in another variation they are reduced to a compact version but not completely removed, and etc. These variations are described in detail in section \ref{sec:ani}. Patterns are extracted from each variation separately using Byte Pair Encoding compression technique. Afterwards, patterns identified from different variations are aggregated and redundant patterns are removed. Each time series is represented by the frequency of these patterns. This pipeline is proposed to keep as much information as possible in the time series representations. 
\begin{figure}
\centering
\includegraphics[width=\columnwidth]{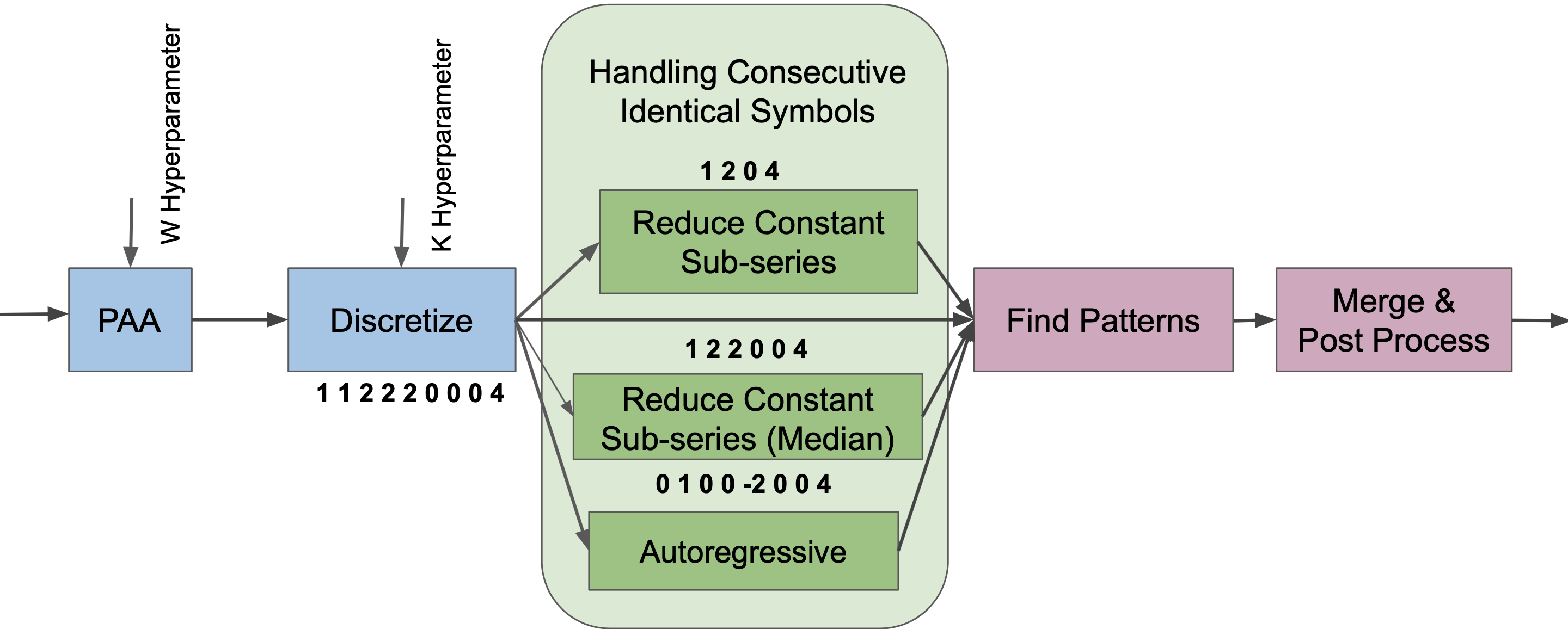}
\caption{Overview of the method. Data is 1.transformed with PAA, 2. discretized, 3. transformed to multiple variations  4. Patterns are extracted from each variation. 5. Features generated from different variations are combined and post-processed to output the final representation.}
\label{fig:method}
\end{figure}
\section{Method}
\label{sec:method}
Figure \ref{fig:method} shows an overview of our proposed method. Data is 1. transformed with PAA, 2. discretized, 3. transformed to multiple variations  4. Patterns are extracted from each variation. 5. Features generated from different variations are combined and post-processed to output the final representation. In the following sections each one of those steps are described in detail. 
We first explain the method for univariate time series, then describe how it can be extended for multivariate series. 
\subsection{PAA transformation}
\label{sec:PAA}
Each time series is first normalized, i.e. zero mean and standard deviation of one. The same normalization step is also applied in the original SAX paper \cite{lin2007experiencing}, as it it shown in \cite{keogh2003need} that time series with different offsets and amplitudes can't be correctly compared with each other.

The normalized time series are then transformed using PAA. PAA reduces the dimension of the time series by splitting them into segments of length $W$ -a hyperparameter of the method- 
and averaging the values in these segments. Since PAA smooths the series, it also helps mitigate the effects of noise and outliers in the data. 
In the left plots of Figure \ref{fig:discrete}, the blue lines (also shown on the right plots) are the PAA-transformed versions of the orange colored time series. 

\subsection{Discretization}
For discretization, first outliers are detected using Inter Quartile Range (IQR) \cite{walfish2006review}. After setting the outliers aside, values are binned (discretized) by equal-width bins. Both outlier detection and the binning are based on the entire dataset, as opposed to each time series independently. Binning the values based on the entire data helps better capture the differences between series. For example if the data is coming from two participants with different levels of physical activity, the discretized data should reflect that (In this case the less active participant will not observe a subset of the symbols that correspond to high levels of activity).  

\begin{figure}
\centering
\includegraphics[width=\columnwidth]{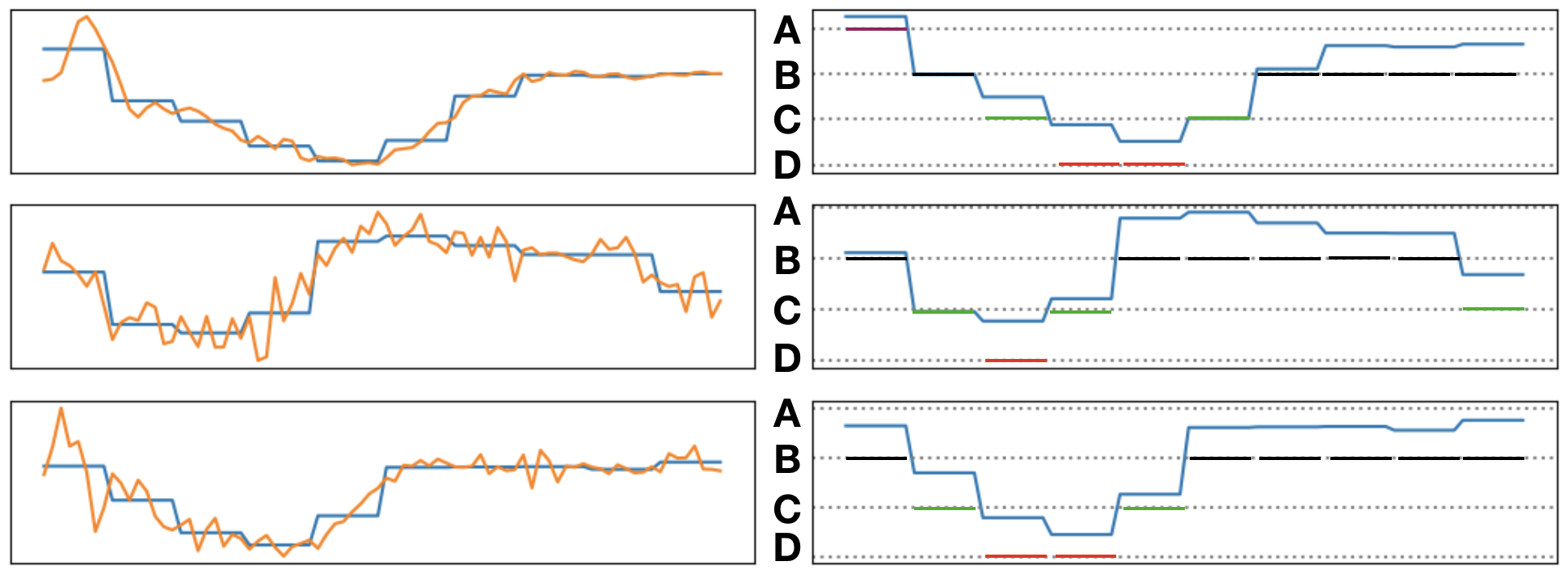}
\caption{Example of time series discretization: The series are first transformed with PAA (on the left), then they are discretized based on equal-width bins (on the right). The discretized version of the first series in this plot is: 'A B C D D C B B B B'. It should be mentioned that PAA actually reduces the length of the series, but here they are plotted the same length with the original signal to show the effect of the transformation.}
\label{fig:discrete}
\end{figure}

Examples of this discretization in shown in the right plots of Figure \ref{fig:discrete}. The PAA-transformed blue series are discretized to 4 bins: A, B, C, and D; each bin/symbol is shown by a different color. The number of bins used for discretization, $K$ is another hyper-parameter of the method. Symbol A is only observed in the first time series. 

\subsection{Identifying Patterns}
We first describe how patterns are extracted from the series, ``Find Pattern`` block in Figure \ref{fig:method}, then discuss its previous step ``Handling Consecutive Identical Symbols`` which addresses how the discretized series are first transformed to different variations before pattern extraction. 

\begin{figure}
\includegraphics[width=0.9\columnwidth]{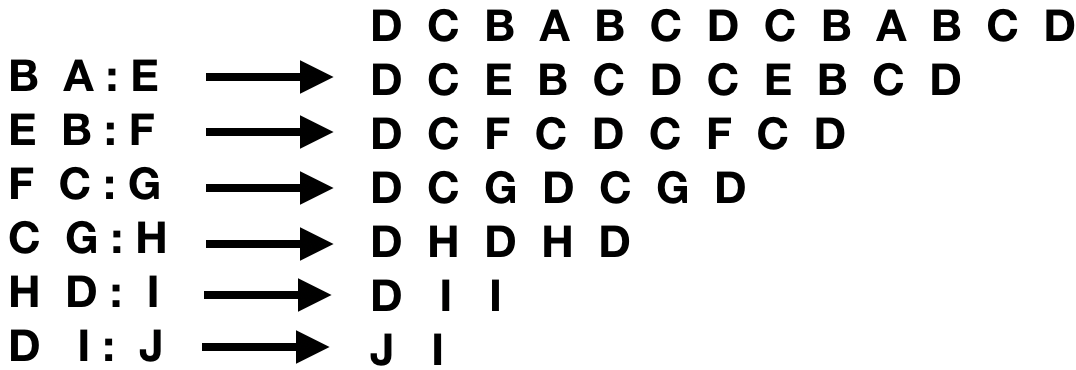}
\caption{Example of BPE on discretized time series: The first line shows the original discretized series. In each iteration the most common pair, is identified and replaced by a new symbol.}
\label{fig:BPI}
\end{figure}

To identify variable length patterns in time series, we use the Byte Pair Encoding (BPE) compression technique \cite{gage1994new}. BPE has been around for a long time, however since it was used in \cite{sennrich2015neural} it received much more recognition. BPE is a compression technique in which the most common pair of consecutive symbols is replaced by a new symbol. In \cite{sennrich2015neural} it was used to address the rare words problem in neural machine translation by subword tokenization. Traditionally, words were considered as tokens in text. With subword tokenization characters are initially considered as tokens and with each iteration two most common tokens are merged together to build a new token. In this way compound words like ``authorship`` could be understood by the model  without having observed it beforehand and by breaking it into subwords ``author-`` and ``-ship``. We use the same approach in identifying patterns in time series. To the best of our knowledge this technique has not yet been used in context of time series. An example of BPE on discretized series is shown in Figure \ref{fig:BPI}. 
In each iteration, the most common pair over all time series in the train set is identified as a new pattern and replaced by a new symbol. For example in Figure \ref{fig:BPI}, in the first iteration, pair B A is identified as a new pattern and is replaced by a new introduced symbol E. The time series shown in the example observes two instances of this pattern, hence, in its representation, the feature corresponding to pattern E has the value 2. This value is later normalized by the length of the series, so that representations of series with different lengths can be compared together.  
With this approach, dimensions of the representation would be number of discretization bins ($K$) + number of identified patterns. For example in Figure \ref{fig:BPI}, the series are discretized to 4 symbols (\textbf{A}, \textbf{B}, \textbf{C}, and \textbf{D}), and 6 new patterns are identified (\textbf{E}, \textbf{F}, \textbf{G}, \textbf{H}, \textbf{I} and \textbf{J}), hence the number of features in this representation is 10. However, this is not always the case. Not all identified patterns are added as new features. They should be present in at least $N\times P, 0<P<1$, where $N$ is the number of series in the dataset, since having sparse features is not beneficial and can lead to overfitting. This constraint also serves as a stopping criteria for finding new patterns. If the overall frequency of the most common pair, $F$, is less than \textbf{MAX($N\times P, T\times U$)}, the loop for finding new patterns is stopped. $T$ is the number of all pairs in the data in the very first iteration 
and $0<U<1$. 
In our experiments, we fix $P$ as $20\%$, and $U$ as $0.001$. We arrived at these numbers empirically based on our experiments on multiple different datasets, although they can also be tuned as extra hyper-parameters. As an example, if data consist of 100 time series with equal lengths of 500, then $N=100$, $T=49900$, and identified patterns must be present in more than 20 time series. The cycle for finding new patterns stops when $F$ becomes less than Max(20, 49.9) = 49.9. Criteria $F < N\times P$ is usually reached in cases where there is a large number of datapoints (time series) in the dataset and criteria $F < T\times U$ is usually reached when the time series are long.

\subsection{Handling Consecutive Identical Symbols}
\label{sec:ani}
When time series are discretized, multiple instances of consecutive identical symbols might be present in the data, for example in the first series in Figure \ref{fig:discrete} ``A B C D D C B B B B``; ``D D`` and ``B B B B`` are instances of consecutive identical symbols.  
As mentioned in the related works section, most similar methods remove consecutive identical patterns and only keep the first one to avoid over counting trivial matches. However, there might be useful information in the number of times symbols are repeated. If we avoid removing consecutive identical symbols, we can extract patterns that capture this useful information, although the method might not be able to extract many other potentially important patterns that become visible only when the repeated symbols are removed. 
For this reason, the series are first transformed to different variations, described below, and patterns are identified from all those variations separately. Afterwards, the identified patterns from different variations are combined together. This is also shown in Figure \ref{fig:method}. 

By using the example in Figure \ref{fig:method}, if the original series is  ``1 1 2 2 2 0 0 0 4``, in the variation RCS (Reduce Constant Sub-series) all repeated symbols are removed and the series is transformed to ``1 2 0 4``. Removing all repeated symbols can reduce the time series significantly, RCSM (Reduce Constant Sub-series (median)) offers a different variation which doesn't reduce the time series as much. In this variation, the median length of constant sub-series is retrieved for all $K$ initial patterns. For a given constant sub-series of pattern i ($s_i$), if the length of $s_i$ is less than median length of all $s_i$, then it's replaced by ``i``, otherwise it's replaced by ``i i``. Based on this variation the example in Figure \ref{fig:method} is transformed to ``1 2 2 0 0 4``, given that median length for each symbol is 2. 

In discretized data, the assumption is that values in different bins have the same distance from each other, however the bins used here are ordered, e.g. in Figure \ref{fig:discrete} bin A is closer to bin B than to bin C. In order to keep this information which is in other words the slope of the time series, the Autoregressive variation is used. By marking the bins as $0 \cdots K-1$, this variation is calculated as follows: $\forall t, y_t = x_t - x_{t-1}$. Going back to the example in Figure \ref{fig:method}, based on this variation the series is transformed to ``0 1 0 0 -2 0 0 4`` (-2 is considered one symbol/pattern).

Consider an example of activity data collected from wearable sensors 
and one pattern (pattern A) that captures exercise activity. Feature corresponding to pattern A, in the original series shows \textbf{how much time overall} each participant exercised  during the study period, in the RCS variation shows \textbf{how many times} each participant exercised, 
the RCSM variation shows  whether a participant had a \textbf{short workout or a long one} compared to all other exercise instances in the dataset, 
and in the Autoregressive variation shows the \textbf{level of the exercise compared to the activity done before}.


\subsection{Post Processing}
Patterns are extracted from each one of the variations described in the previous section separately. Afterwards, all the identified patterns from different variations are combined together. Some patterns might exist in more than one variation, in order to avoid having redundant features corresponding to a single pattern, highly correlated features,  with more than 0.95 correlation, are removed. Features with 0 variance are also removed because they do not contain useful information in representing the series. 

\subsection{Extension to Multivariate Series}
\label{sec:multi}
This process can also be extended to multivariate time series. We use the approach SAX-ZSCORE proposed in \cite{mohammad2014robust}, to extend this method.

The first step discussed in Section \ref{sec:PAA} for extracting patterns is normalizing the series. In the case of multivariate series, an extended multidimensional version of the zscore normalization is used. Assuming that the covariance matrix of the multivariate time series $x$, is $C$ and its multi-dimensional mean is $\mu$, the normalized time series is:  
\begin{equation}
\hat{x} = C^{-1/2}(x - \mu).
\end{equation}
Where $C^{-1/2}$ is the inverse of Cholesky decomposition. 
Afterwards, $L_2$ norm of $\hat{x}$ is calculated which outputs a one dimensional series. These one dimensional time series are then fed into the method the same way as before and patterns are extracted from them. 
\subsection{Classification/Regression}
Since the representations are unsupervised, any classifier/regressor can be used for the down-stream task. In our experiments, Support Vector Machine (SVM) with RBF kernel generally performed better than other methods. This method has two main hyper-parameters, $C$ and $\gamma$. $C$ is the regularization parameter and $\gamma$ is the kernel coefficient.
\subsection{Hyper-parameter Tuning}
For optimizing hyper-parameters, Hyperopt is used \cite{bergstra2013making}. Hyperopt is a Python library with Bayesian optimization for parameter tuning. It uses the results of past evaluations to learn a probabilistic model for the objective function, and chooses the next set of values for hyper-parameters based on the learned model. 
Our proposed approach contains four parameters that are tuned using Hyperopt. These parameters are listed below:
\begin{itemize}
\item $K$: Number of bins in discretization, $2\leq K\leq 100$
\item $W$: Window size in PAA transformation, $1\leq W\leq 15$
\item $C$ \& $\gamma$: SVM parameters
\end{itemize}

\section{Data}
\label{sec:data}
The data used in this paper comes from a real-world study called TILES (Tracking Individual Performance with Sensors) \cite{mundnich2020tiles} which studies the relationship between physical activity and physiological states. The data was collected during a 10-week long study that recruited 212 hospital workers.  
Participants' bio-behavioral data was captured via different wearable devices during the study. In this paper we focus on data collected from Fitbit wristbands.
The Fitbit wristband captures heart rate and step count. 
For this work, the data was pre-processed and averaged to 5 minute intervals. This yields heart rate and step count time series of length 288 (24 hours/ 5 minutes) for each day. 

The studies also administered surveys to collect self-assessments of individual participant's stress, sleep, job performance, organizational behavior, and other personality constructs. The goal is to use time series collected from fitbit to estimate these constructs, meaning the constructs are used as targets for time series classification/regression problems. The labels are considered in two groups: pre-study surveys and daily surveys. 
Before the start of the study, participants were asked to fill out surveys which assessed their job performance, cognitive ability, and health. We call these constructs IGTB (Initial Ground Truth Battery). For IGTB constructs, there is one label per participant. Participants also answered daily surveys, hereafter called MGT (Mobile ground truth). 
For MGT constructs there is one label per day for each participant. A subset of these constructs are described below. 

 Big 5 personality traits \cite{goldberg1992development} is among the most commonly used models of personality in psychology. It consists of 5 categories: Openness (OPE), Conscientiousness (CON), Extraversion (EXT), Agreeableness (AGR), and  Neuroticism (NEU). A big 5 personality test assigns a number to each one of these 5 categories based on the answered questions. Each category represents a range between two extremes, for example extreme extraversion and extreme introversion. This test is taken from participants in both pre-study surveys and daily surveys, therefore these 5 labels are in both IGTB and MGT constructs. Constructs also include self-assessments of job performance (Individual Task Proficiency(ITP) and In-Role Behaviors(IRB)), alcohol and tobacco use, sleep quality, stress, anxiety, positive and negative affect (Pos/Neg-Affect), etc. 
 
 Most of the constructs have numerical values and are treated as regression problems, however there are a couple of categorical constructs in MGT, Atypical, Work, Location, Activity, and Interaction. In Atypical, participants report if they had experienced, or anticipate experiencing an atypical event; Work: whether that day was a work-day; Location: the location where participants spent most of the day in, and etc. These constructs are treated as classification problems. 
 
Other IGTB and MGT constructs not discussed in text, are listed in Table \ref{tab:construct}. For more information on the constructs and their survey references please refer to the original paper \cite{mundnich2020tiles}.

 \begin{table}
 \begin{center}
\begin{tabular}{ |c|c| } 
 \hline
 Construct & Abbreviation   \\ 
 \hline\hline
  Counterproductive Work Behavior (ID) & IOD-ID   \\ 
  \hline
  Counterproductive Work Behavior (OD) & IOD-OD  \\ 
  \hline
 Organizational Citizenship Behavior & OCB \\ 
  \hline
Cognitive ability (Shipley Abstraction) & Ship-Abs  \\ 
\hline
 Cognitive ability (Shipley Vocabulary) & Ship-Voc  \\ 
  \hline
State Trait Anxiety Inventory & STAI  \\ 
\hline
Alcohol Use Disorders Identification Test & AUDIT  \\ 
\hline
 Tobacco units used in past week & Gats-Quant  \\ 
\hline
 International Physical Activity Questionnaire & IPAQ  \\ 
\hline
Pittsburgh Sleep Quality Index & PSQI  \\ 
\hline
Counterproductive Work Behavior & CWB  \\ 
\hline
\end{tabular}
\caption{Remaining IGTB \& MGT constructs not discussed in text.} 
\label{tab:construct}
\end{center}
\end{table}

Since participants have different compliance rates in wearing the sensors, different amount of data is collected from them, i.e. each participant wears the sensor for different number of days and different number of hours in each day. For the MGT labels, the length of each data-point, time series corresponding to one day, is kept fixed to 288 time steps and the missing data is filled with 0. The days where the collected data overall was less that 6 hours (72 data points) were removed from the dataset. For IGTB constructs each data point corresponds to all data collected from one participant over the study period, which consists of different number of days. For these constructs, data points (corresponding to participants) with less than 7 days worth of data were removed. Data from both IGTB and MGT constructs have noise and missing values, IGTB labels have time series with variable length (different number of days), while in MGT, time series all have the same length. 

In daily surveys, MGT, most of the times participants don't answer all of the questions, therefore, for each MGT construct there are different number of answers (labels). In general there are 12391 number of days collected from 205 participants in the dataset. Number of labels for different constructs ranges between 725 and 10220, with average of 5374 data-points. For IGTB constructs, overall there are 206 participants with more than 6 days worth of data. Number of data-points for different IGTB constructs ranges between 198 and 206 with average number of 205. In IGTB dataset number of days -length of the time series- ranges between 7 and 85 with 61 average number of days for each participant. 

\section{Results}
As described in Section \ref{sec:data}, Fitbit wristbands collect steps and heart-rates of participants. In order to estimate IGTB/MGT constructs from this data, patterns are extracted from each modality separately (heart rate and steps), and the resulting representations are concatenated together. Table \ref{tab:IGTB} shows the results of our proposed method Pattern Discovery with Byte Pair Encoding (PD-BPE),  time series embedding proposed in \cite{franceschi2019unsupervised} shown here as TS-Embed, and Rocket \cite{dempster2019rocket}, for IGTB constructs. For both TS-Embed and Rocket the implementation and hyper-parameters suggested by their corresponding authors were used. The results are reported in RMSE (Root Mean Squared Error) and the \#Patterns column shows the number of patterns identified in the dataset by PD-BPE from both heart rate and step count series. The results are based on 5-fold cross validation and hyper-parameter optimization is done using nested cross-validation. The same folds were used in all three models. The best performing model's results are highlighted in bold. Based on the results in Table \ref{tab:IGTB}, PD-BPE outperforms other methods in 12 out of 19 IGTB constructs. 

\begin{table}
\begin{center}
\begin{tabular}{ |c||c|c||c|c| } 
 \hline
 IGTB & PD-BPE & \small{\#Patterns} & TS-Embed &  Rocket  \\ 
 \hline\hline
 ITP & 0.530 & 624 & 0.531 & \textbf{0.527} \\ 
  \hline
 IRB & 4.417 & 536 & 4.472 & \textbf{4.395} \\ 
  \hline
 IOD-ID & \textbf{4.986} & 623 & 5.251 & 5.114 \\ 
  \hline
 IOD-OD & \textbf{6.695} & 523 & 6.916 & 6.738 \\ 
  \hline
 OCB & 12.250 & 640 & 12.291 & \textbf{12.237} \\ 
  \hline
 Ship-Abs & 3.929 & 491 & 3.934 & \textbf{3.920} \\ 
  \hline
 Ship-Voc & \textbf{4.790} & 421 & 5.017 & 5.102 \\ 
  \hline
 NEU & \textbf{0.710} & 639 & 0.730 & 0.724 \\ 
  \hline
 CON & \textbf{0.603} & 623 & 0.629 & 0.622 \\ 
  \hline
 EXT & 0.648 & 511 & 0.643 & \textbf{0.642} \\ 
  \hline
 AGR & \textbf{0.480} & 628 & 0.485 & 0.485 \\ 
  \hline
 OPE & \textbf{0.569} & 637 & 0.575 & 0.584 \\ 
  \hline
 Pos-Affect & \textbf{6.628} & 735 & 6.670 & 6.671 \\ 
  \hline
 Neg-Affect & \textbf{5.187} & 486 & 5.437 & 5.281 \\ 
  \hline
 STAI & 8.813 & 662 & 8.912 & \textbf{8.802} \\ 
  \hline
 AUDIT & \textbf{2.278} & 529 & 2.301 & 2.309 \\ 
  \hline
 Gats-Quant & \textbf{3.977} & 490 & 4.040 & 4.247 \\ 
  \hline
 IPAQ & 17117 & 311 & 17133 & \textbf{16096} \\ 
  \hline
 PSQI & \textbf{2.359} & 730 & 2.374 & 2.384 \\ 
  \hline
\end{tabular}
\caption{Results for numerical IGTB constructs, reported in RMSE}
\label{tab:IGTB}
\end{center}
\end{table}

In MGT constructs, multiple data points belong to the same participant. 
Here the same approach as \cite{burghardt2020having} is used to control for subject-specific differences in behavior. For each participant, average of their representations across different days in computed, the centroid representation of that participant. Afterwards, representations learned for each day is combined (concatenated) with centroid representation of that individual . Table \ref{tab:MGT} shows the results for numerical MGT constructs. For all three methods, daily representations are concatenated with participants' centroid representation. Based on the results in Table \ref{tab:MGT} PD-BPE outperforms other methods in 15 out of 17 MGT constructs. Table \ref{tab:class} shows the results for categorical MGT constructs. The results are reported in accuracy, except for Atypical construct which is reported in AUC-ROC (Area Under the Receiver Operating Characteristic Curve) because of its imbalance. 91\% of the datapoints in Atypical construct are 0 (typical). Based on the results in Table \ref{tab:class} PD-BPE outperforms other methods in 2 out of 5 MGT constructs. 

Heart rate and step count time series can be considered together as a multivariate time series. With the extension described in Section \ref{sec:multi}, patterns are extracted in multivariate setting and results for a subset of MGT constructs are reported in Table \ref{tab:multivar}. It is shown that PD-BPE outperforms competing methods in 5 out of 5 MGT constructs. 

Based on the results reported in Tables \ref{tab:IGTB}, \ref{tab:MGT}, PD-BPE  outperforms both methods in regression problems. This improvement is kept intact when extending to multivariate time series (Table \ref{tab:multivar}). However, in categorical constructs Rocket outperforms PD-BPE in 3 out of 5 constructs (Table \ref{tab:class}). 

\subsection{Computation Speed}
Although it's worth noting that PD-BPE generates interpretable and far fewer features than Rocket. Number of PD-BPE features for MGT constructs are twice the \#Patterns column, because of combining the participant representations with daily representations. Based on constructs of all three tables \ref{tab:IGTB}, \ref{tab:MGT} and \ref{tab:class}, PD-BPE has an average of 737 features with standard deviation of 283. Whereas number of Rocket and TS-Embed features are 80000 and 5120 respectively. The proposed number of dimensions for Rocket and TS-Embed representations are 20000 and 1280 respectively, here because of combing representation of heart rate and step count along with daily and participant representations, these numbers are multiplied by 4. 

The dimension of feature space also reflects in the speed of computations. On a cluster with Intel Xeon E5-2650 v4 CPUs; Rocket, known for its fast speed, takes \textbf{336} seconds to generate and apply kernels to the MGT dataset and \textbf{137} seconds on average to train linear classifiers (10 fold) on categorical constructs. On the same machine, PD-BPE takes \textbf{32} seconds and \textbf{145} seconds to extract patterns and fit non-linear classifiers respectively. TS-Embed trains the whole neural network and it's computation speed is not comparable with Rocket and PD-BPE. 
\subsection{interpretability}
\begin{figure}
\centering
\includegraphics[width=\columnwidth]{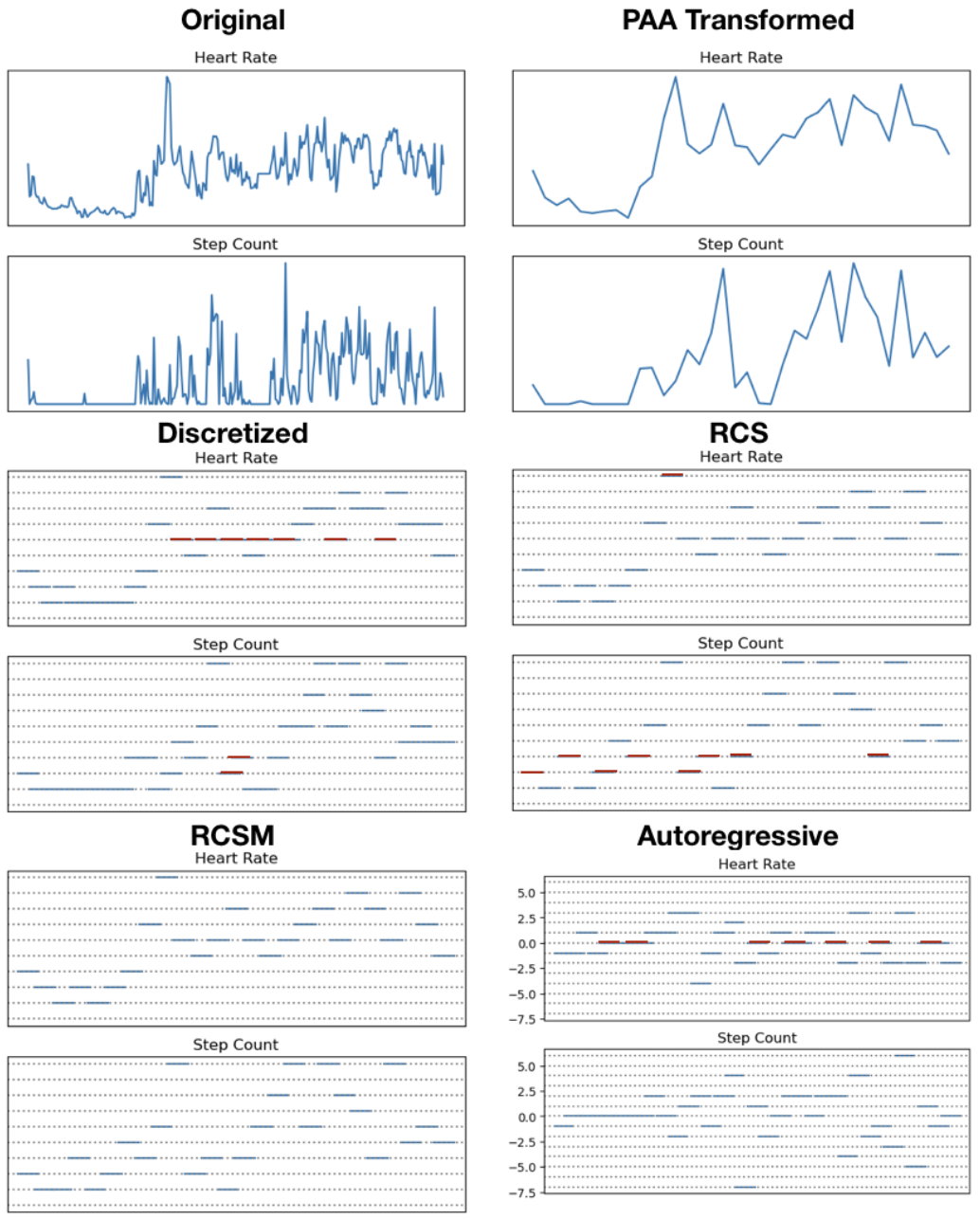}
\caption{A sample datapoint for Atypical event construct. Top left plots show the original heart rate and step count. Top right plots show the PAA transformed data with window size $W$ = 8. In middle left, the data is discretized into $K$ = 10 bins. Middle right shows the RCS variation. Bottom left is the RCS Median variation. And finally bottom right is the autoregressive variation. 10 most important patterns for classifying Atypical events, based on F-values from ANOVA test, are highlighted in red. }
\label{fig:viz}
\end{figure}

Since PD-BPE is interpretable, it can offer a better understanding of both the problem and the representations generated. As an example we look at the patterns identified for Atypical construct. 
Based on F-values from ANOVA test, the most important features for identifying atypical days, correspond to patterns that capture long inactivity intervals. For example one of the most important patterns is a 5 hour long pattern, extracted from the Autoregressive variation of step count modality, in which the step count remains almost constant. Further analysis shows that this pattern is usually observed during the night, which means it is probably capturing sleep time. The frequency of observing this pattern is much higher in typical vs atypical days, forming the hypothesis that participants experiencing atypical days, are most likely to lose sleep and capturing sleep time of participants can help identify atypical days. Figure \ref{fig:viz} shows a sample of an atypical event. The most important patterns, in different modalities and variations, are highlighted in red.  
\section{Conclusion} 
In this work, an interpretable and scalable method for learning representations from time series is proposed. This method extracts variable length patterns from data by discretizing the time series and using BPE compression technique. Using data collected from wearable sensors, it was shown that for regression tasks, the proposed approach beats state-of-the-art methods and for classification tasks, it gives comparable results. 

Although this method was proposed for dealing with unclean and noisy datasets -such as data collected from wearable sensors- it could also be applied to any other time series dataset. One possible future work direction is to evaluate results of PD-BPE method on cleaner and more structured datasets. Also this method could be modified to perform better for classification problems for example by identifying patterns in a supervised manner. 
Another possible direction for improving the method further is to use Discrete Fourier Transform on windows instead of PAA, similar to BOSS method \cite{schafer2015boss}. Using DFT will decrease the interpretability of the method to some degree but will most likely improve the performance, as it has been shown in \cite{schafer2015boss}.
\begin{table}
\begin{center}
\begin{tabular}{ |c||c|c||c|c| } 
 \hline
 MGT & PD-BPE & \small{\#Patterns} & TS-Embed &  Rocket  \\ 
 \hline\hline
 ITP & \textbf{0.562} & 475 & 0.654 & 0.606 \\ 
  \hline
 IRB & \textbf{5.156} & 375 & 5.998 & 5.665 \\ 
  \hline
 OCB & \textbf{0.813} & 511 & 0.875 & 0.847 \\ 
  \hline
 CWB & \textbf{1.202} & 574 & 1.388 & 1.298 \\ 
  \hline
 NEU & \textbf{0.865} & 558 & 1.022 & 0.988 \\ 
  \hline
 CON & \textbf{0.689} & 374 & 0.780 & 0.765 \\ 
  \hline
 EXT & \textbf{0.790} & 550 & 0.929 & 0.893 \\ 
  \hline
 AGR & \textbf{0.737} & 484 & 0.877 & 0.807 \\ 
  \hline
 OPE & \textbf{0.732} & 409 & 0.858 & 0.828 \\ 
  \hline
 Pos-Affect & \textbf{3.466} & 599 & 4.322 & 3.555 \\ 
  \hline
 Neg-Affect & \textbf{2.133} & 378 & 2.370 & 2.167 \\ 
  \hline
 Anxiety & \textbf{0.687} & 408 & 0.735 & 0.691\\ 
  \hline
 Stress & \textbf{0.830} & 296 & 0.895 & 0.835\\ 
  \hline
 Alcohol & \textbf{1.200} & 259  & 1.233 & 1.219\\ 
  \hline
Tobacco & \textbf{0.564} & 689  & 0.880 & 0.605\\ 
\hline
Exercise & 632.2 & 311 & 659.8 & \textbf{497.2}\\ 
  \hline
Sleep & 1.750 & 236 & 1.789 & \textbf{1.737}\\ 
\hline
\end{tabular}
\end{center}
\caption{Results for numerical MGT constructs, reported in RMSE}
\label{tab:MGT}
\end{table}
\begin{table}
\begin{center}
\begin{tabular}{ |c||c|c||c|c| } 
 \hline
 MGT  & PD-BPE & \small{\#Patterns} & TS-Embed &  Rocket  \\ 
 \hline\hline
 Atypical & \textbf{0.774} & 460 & 0.673 & 0.746 \\ 
  \hline
 Work & 0.831 & 183 & 0.869 & \textbf{0.888} \\ 
  \hline
 Location & 0.611 & 305 & 0.610 & \textbf{0.621} \\ 
  \hline
 Activity & 0.401 & 399 & 0.385 & \textbf{0.403} \\ 
  \hline
 Interaction & \textbf{0.603} & 855 & 0.576 & 0.598 \\ 
  \hline
\end{tabular}
\end{center}
\caption{Results for categorical MGT constructs, reported in accuracy except for Atypical construct, which is reported in AUC ROC}
\label{tab:class}
\end{table}
\begin{table}
\begin{center}
\begin{tabular}{ |c||c|c||c|c| } 
 \hline
 MGT & PD-BPE & \small{\#Patterns} & TS-Embed &  Rocket  \\ 
 \hline\hline
 ITP & \textbf{0.567}  & 365 & 0.649  & 0.592 \\ 
  \hline
 IRB & \textbf{5.220} & 375 & 5.908 & 5.543 \\ 
  \hline
 OCB & \textbf{0.796} & 289 & 0.881 & 0.835 \\ 
  \hline
 CWB & \textbf{1.204} & 360 & 1.389 & 1.280 \\ 
  \hline
 NEU & \textbf{0.861} & 314 & 1.021 & 0.909 \\ 
  \hline
\end{tabular}
\end{center}
\caption{Results for a subset of numerical MGT constructs with multivariate data, reported in RMSE.}
\label{tab:multivar}
\end{table}

\section{Acknowledgements}
The research was supported by the Office of the Director of National
Intelligence (ODNI), Intelligence Advanced Research Projects Activity
(IARPA), via IARPA Contract No 2017-17042800005.

\bibliographystyle{ACM-Reference-Format}
\bibliography{sample-base}

\end{document}